    \documentclass[runningheads]{llncs}

\usepackage[T1]{fontenc}
\usepackage{graphicx}
\usepackage{amsmath,amssymb,amsfonts}
\usepackage{algorithm}
\usepackage{algpseudocode}
\usepackage{textcomp}
\usepackage{xcolor}
\usepackage{tikz}
\usepackage{pgfplots}
\pgfplotsset{compat=1.18}
\usepgfplotslibrary{statistics}
\usepackage{url}
\usepackage{booktabs}
\usepackage{multirow}
\usepackage{subcaption}
\usepackage{enumitem}
\usepackage{balance}
\usepackage{colortbl}
\usepackage{xspace}
\usepackage{orcidlink}

\newcommand{\minisection}[1]{\paragraph{#1}}

\providecommand{\SetKwInOut}[2]{}  
\providecommand{\SetKwFunction}[2]{}  

\providecommand{\While}[2]{\textbf{while} #1 \textbf{do}\\#2\\\textbf{end while}}
\providecommand{\If}[2]{\textbf{if} #1 \textbf{then}\\#2\\\textbf{end if}}

\newcommand{\OurMethod}{\texttt{EvolveGen}\xspace}
\newcommand{\HLSModel}{\texttt{HLS2Model}\xspace}

    \usepackage{booktabs}
    \usepackage{bbding} 

\ifdefined\ACM
\else\ifdefined\LNCS
\else
    \usepackage{cite}
\fi\fi

\definecolor{lightblue}{RGB}{223, 235, 247}

\def\BibTeX{{\rm B\kern-.05em{\sc i\kern-.025em b}\kern-.08em
    T\kern-.1667em\lower.7ex\hbox{E}\kern-.125emX}}
\newlist{myitemize}{itemize}{1}
\setlist[myitemize]{
  label=\textbullet, 
}

\usepackage{etoolbox}
\makeatletter
\patchcmd{\@makecaption}
  {\scshape}
  {}
  {}
  {}
\makeatother

\def\BibTeX{{\rm B\kern-.05em{\sc i\kern-.025em b}\kern-.08em
    T\kern-.1667em\lower.7ex\hbox{E}\kern-.125emX}}

\makeatletter
\newcommand\printfnsymbol[1]{%
  \textsuperscript{\@fnsymbol{#1}}%
}
\renewcommand*{\@fnsymbol}[1]{\ensuremath{\ifcase#1\or *\or \dagger\or \ddagger\or \mathsection\or \mathparagraph\or \|\or **\or \dagger\dagger \or \ddagger\ddagger \else\@ctrerr\fi}}
\makeatother

\begin{document}


\title{\OurMethod{}: Algorithmic Level Hardware Model Checking Benchmark Generation through
Reinforcement Learning}
\titlerunning{\OurMethod{}: RL-Based Hardware MC Benchmarks Generation}




%



\author{Guangyu~Hu\orcidlink{0000-0001-5077-8361}\thanks{Both authors contributed equally to this work.}\inst{1} \and
Xiaofeng~Zhou\orcidlink{0000-0001-5878-3683}\printfnsymbol{1}\inst{1} \and
Wei~Zhang\orcidlink{0000-0002-7622-6714}\textsuperscript{(\Envelope)}\inst{1} \and
Hongce~Zhang\orcidlink{0000-0003-4001-264X}\textsuperscript{(\Envelope)}\inst{2}}
\authorrunning{G. Hu and X. Zhou et al.}
\institute{The Hong Kong University of Science and Technology\\
\email{\{ghuae,xzhoubu\}@connect.ust.hk, wei.zhang@ust.hk} \and
The Hong Kong University of Science and Technology (Guangzhou)\\
\email{hongcezh@hkust-gz.edu.cn}
}

\ifdefined\ACM
    \begin{abstract}
%
%



The advancement of hardware model checking is critically dependent on high-quality benchmarks. However, the community faces a significant benchmark gap: existing benchmark suites are limited in number, primarily provided at representations like BTOR2 without access to the originating register-transfer-level (RTL) designs, and skewed toward extreme difficulty levels, where instances are either trivial or intractable. These limitations not only impede the rigorous evaluation of new verification techniques but also encourage the overfitting of solver heuristics to a narrow set of problems.

To address this challenge, we introduce \OurMethod{}, a novel framework that generates hardware model checking benchmarks by combining reinforcement learning (RL) with high-level synthesis (HLS). Our approach operates at an algorithmic level of abstraction, where an RL agent learns to construct computation graphs. By compiling these graphs with different synthesis directives, we generate pairs of functionally equivalent but structurally distinct hardware designs,
thereby inducing challenging model checking instances.
The solver's execution time serves as a reward signal, guiding the agent to autonomously discover and generate ``small-but-hard'' instances that target solver-specific weaknesses. Our experiments show that \OurMethod{} efficiently creates a diverse benchmark set in various standard formats (e.g., AIGER, BTOR2), effectively revealing performance bottlenecks in state-of-the-art model checkers.

\end{abstract}

    \maketitle
\else
    \maketitle
    
\fi
\section{Introduction}
\label{sec:intro}

The progress of hardware model checking, a cornerstone of hardware formal verification~\cite{Kern1999}, is fundamentally tied to the quality and diversity of available benchmarks. However, the community faces a chronic \textit{shortage}: despite annual calls for benchmarks from venues like the hardware model checking competition (HWMCC)~\cite{biere2024hwmcc}, obtaining authentic open-source hardware designs remains a major challenge. Consequently, a significant portion of widely-used suites originates from software or protocol verification, with only about 67\% of recent HWMCC instances providing synthesizable register-transfer-level (RTL) descriptions~\cite{biere2024hwmcc}. This scarcity is compounded by a severe distribution problem. Existing suites are often polarized, filled with instances that are either trivially easy or extremely hard. Furthermore, as model checkers advance, previously challenging benchmarks may become obsolete, creating a persistent lack of the ``solver-specific-challenging'' instances crucial for robust evaluation. This environment fosters a damaging cycle of ``over-training,'' where solver heuristics are excessively tuned to a narrow, potentially biased, and sometimes even ``badly encoded'' set of problems.
It is beneficial and essential to have a 
 systematic approach to generate diverse, structurally complex, and algorithmically challenging hardware model checking benchmarks. 
\paragraph{Related work.} 
There have been several prior works that automatically generate circuits to test verification tools, such as \textbf{AIGFuzz}~\cite{Zhang2021}, \textbf{AIGen}~\cite{Jacobs2021}, and \textbf{FuzzBtor2}~\cite{xiao2023fuzzbtor2}.
Machine learning techniques, such as reinforcement learning (RL), have also been integrated, for example, in \textbf{BanditFuzz}~\cite{scott2020banditfuzz}.
While prior works generate syntactic mutants to better trigger tool bugs, they cannot generate benchmarks whose difficulty stems from complex semantic and structural properties, which is crucial to stress the performance of hardware model checkers.


\paragraph{Contributions.} This paper introduces \OurMethod{}, a novel framework that leverages RL and high-level synthesis (HLS) to generate benchmarks from a high-level algorithmic abstraction. 
The RL agent learns to construct a computation graph representing an algorithm. This graph is then compiled into two functionally equivalent but distinct hardware designs using varied HLS directives. 
A reward signal, based on a model checker's solving time, guides the agent to autonomously discover ``small-but-hard'' problems. 
These benchmarks can be generated in large scale, and they also expose structural bottlenecks that cause solver heuristics to fail.
This is the first framework to generate hardware benchmarks at an algorithmic level. Our evaluation shows that these benchmarks reveal performance differences among the state-of-the-art model checkers.
%
\paragraph{Structure of the Paper.} The remainder of this paper is structured as follows. Section~\ref{sec:preliminary} provides the necessary background. Section~\ref{sec:motivation} presents our motivating example and key observations. Section~\ref{sec:method} details the proposed \OurMethod{} framework. Section~\ref{sec:experiment} presents the experimental results. Finally, Section~\ref{sec:conclusion} concludes the paper.

%
%
\section{Preliminaries}
\label{sec:preliminary}
%
%

\paragraph{Hardware Model Checking.}
Hardware model checking automatically verifies whether a hardware design satisfies a given set of properties~\cite{clarke1999model}. The design is typically represented as a state transition system, 
while properties can be specified using a certain temporal logic~\cite{pnueli1977temporal}. A model checker explores the state space of the design to either prove that a property holds for all possible behaviors or provide a counterexample trace that demonstrates a violation. The performance and capabilities of model checkers are typically evaluated using benchmark suites like the HWMCC benchmark set, making the quality and diversity of these benchmarks critical for advancing the field.
\paragraph{High-Level Synthesis.}
HLS is a design methodology that automates the creation of hardware from a high-level algorithmic description, typically written in languages like C, C++, or SystemC~\cite{ieee2011systemc}. An HLS compiler analyzes the high-level code and synthesizes it into an RTL hardware implementation (e.g., in Verilog or VHDL). A key feature of HLS is the use of compiler directives, or \textit{pragmas}, which allow designers to guide the synthesis process. For example, pragmas can specify how loops should be unrolled or pipelined and how arrays are mapped to memory. Two functionally identical C++ programs can be synthesized into vastly different hardware implementations by applying different pragmas, a characteristic that our framework leverages to create equivalence checking challenges.
\paragraph{Reinforcement Learning.}
Reinforcement Learning (RL) is a machine learning approach where an agent learns to make optimal decisions by interacting with an environment to maximize a cumulative reward signal~\cite{Sutton2018}. RL problems are typically formalized as Markov Decision Processes (MDPs)~\cite{puterman2014markov} involving complex state transitions. The Multi-Armed Bandit (MAB) problem~\cite{gittins2011multi} represents a fundamental subclass of RL characterized by a single state and a finite set of available actions, $\mathcal{A}$. In the MAB setting, the agent repeatedly selects an action $a \in \mathcal{A}$ (often referred to as pulling an ``arm'') and observes a stochastic reward drawn from an unknown underlying distribution. The objective is to maximize the total reward over time by balancing \textit{exploration} (trying those less-frequently-tested arms) and \textit{exploitation} (choosing the arm estimated to be the best).

A powerful algorithm for solving MAB problems is \textit{Thompson Sampling}~\cite{agrawal2012analysis}, which employs a Bayesian approach for decision-making. In scenarios where rewards are binary (Bernoulli distributed), Thompson Sampling maintains a Beta distribution, $\text{Beta}(\alpha, \beta)$, for each arm to model the probability of receiving a reward. The shape parameters $\alpha$ and $\beta$ correspond to the accumulated counts of successes (reward $=1$) and failures (reward $=0$), respectively. In each decision step, the agent samples a value from the Beta distribution of every arm and greedily selects the action corresponding to the highest sampled value. Upon observing the actual reward, the agent updates the $\alpha$ or $\beta$ parameter of the chosen arm, thereby refining its belief model.

\section{Motivation}
\label{sec:motivation}
\minisection{Analysis of Existing Benchmarks.}

\begin{figure*}[ht]
    \centering
    \begin{subfigure}[b]{0.48\textwidth}
        \centering
        \includegraphics[width=\textwidth]{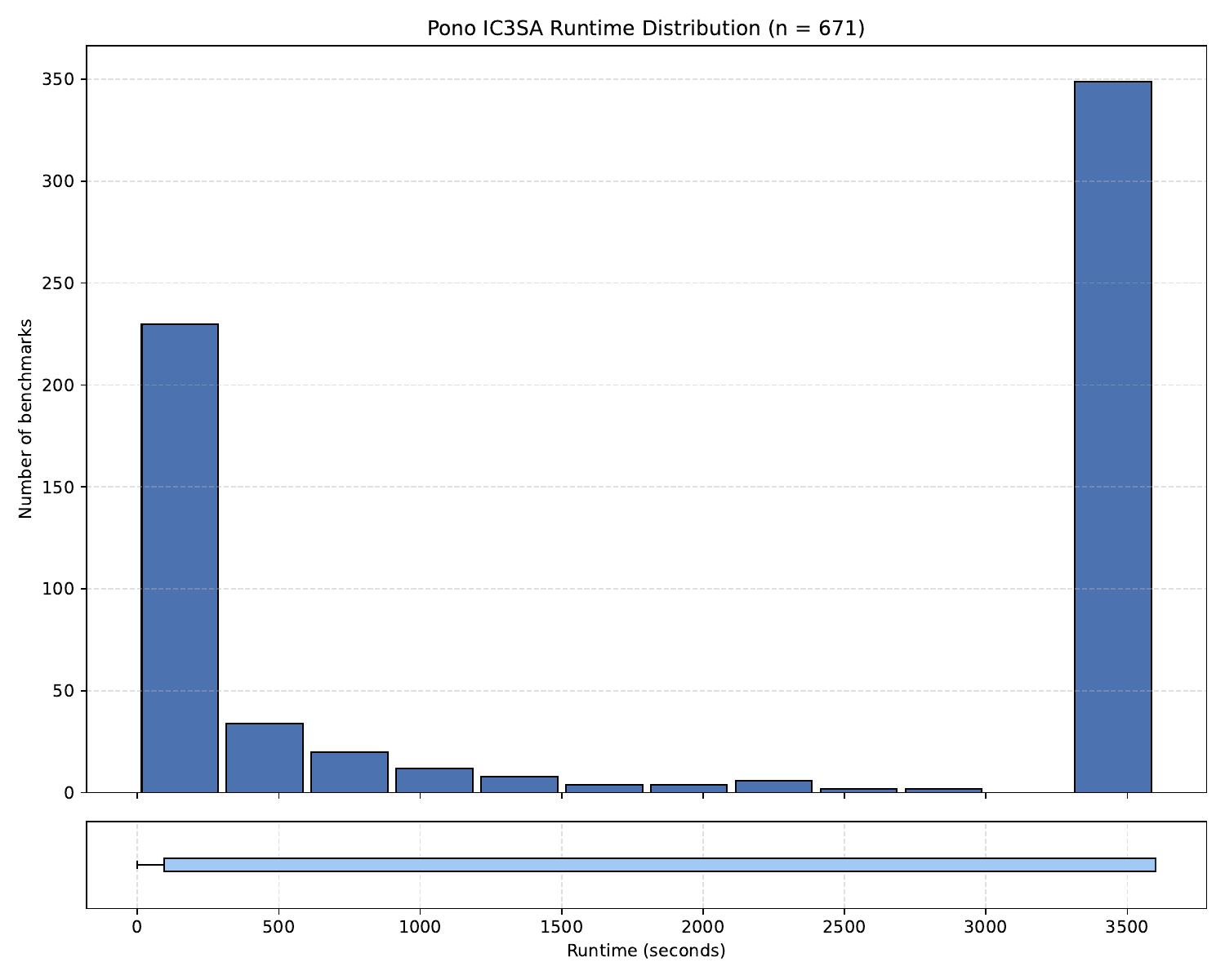}
        \caption{Difficulty spectrum (solving time distribution by Pono)}
        \label{fig:motivation-a}
    \end{subfigure}
    \hfill
    \begin{subfigure}[b]{0.48\textwidth}
        \centering
        \includegraphics[width=\textwidth]{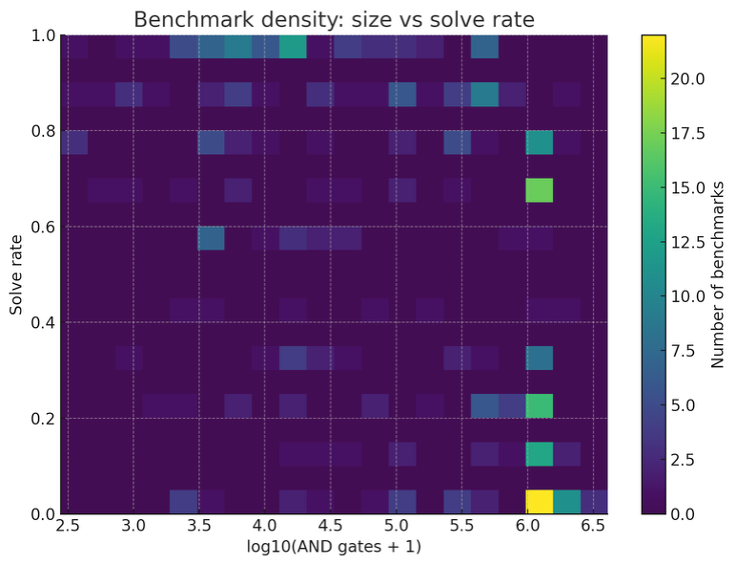}
        \caption{Benchmark density heatmap (size vs. time by all solvers that participated in HWMCC'24)}
        \label{fig:motivation-e}
    \end{subfigure}
    \caption{Analysis of the HWMCC'24 benchmark suite. (a) The difficulty spectrum is highly polarized, with a scarcity of mid-difficulty instances (600s-3600s). Solving time is measured on Pono. (b) Benchmark difficulty is strongly correlated with circuit size, indicating scale-driven difficulty.}
    \label{fig:motivation_summary}
\end{figure*}

To advance hardware model checking, the community requires benchmarks that are not only challenging but also diverse in the nature of their difficulty. However, our analysis of the widely-used HWMCC'24 dataset reveals two critical limitations in the current benchmark landscape, as summarized in Fig.~\ref{fig:motivation_summary}.

First, the difficulty spectrum is poorly distributed. As shown in Fig.~\ref{fig:motivation-a}, the dataset is dominated by instances that are either trivially easy (solved in under 100 seconds) or intractably hard (unsolved after one hour). The crucial ``mid-difficulty'' range, which is most effective for distinguishing the performance of competitive solvers, is sparsely populated. 


Second, benchmark difficulty is primarily driven by the scale of the problem, rather than the underlying structure.
The heatmap in Fig.~\ref{fig:motivation-e} reveals a notable correlation between circuit size (number of AIG gates) and the solving time. 
%
There is a severe lack of ``small-but-hard'' circuits, 
which are ideal for diagnosing solver behavior, 
and challenge a solver's core \textit{inductive reasoning} capabilities. 
%
When a solver fails on a small but hard instance, it points to a fundamental weakness in its reasoning logic, which might be obscured in a larger design. This observation highlights a clear need for benchmarks that are structurally complex and can rigorously test the inductive power of modern solvers.

\begin{figure}[htbp]
    \centering
    \includegraphics[width=0.95\textwidth]{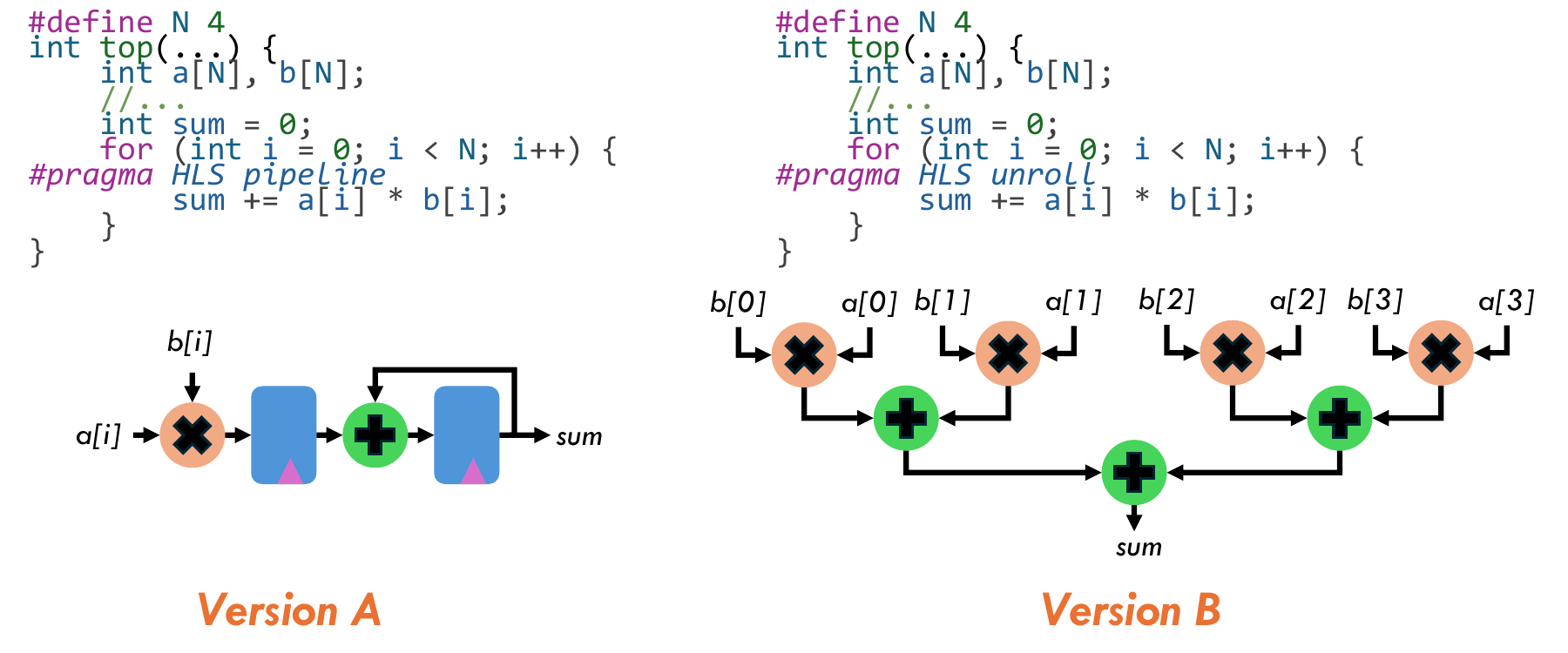}
    \caption{Two functionally equivalent C++ implementations of a multiply-accumulate loop, differing only by HLS pragmas.  Version A is pipelined, while Version B is fully unrolled, leading to vastly different hardware implementations.}
    \label{fig:motivation_placeholder}
\end{figure}

%
\minisection{A Motivating Example.}
Consider the simple C++ function shown in Fig.~\ref{fig:motivation_placeholder}, which performs a multiply-accumulate operation within a loop. From this single high-level description, we can generate two functionally identical hardware implementations using HLS by applying different optimization directives (pragmas).\looseness=-1

Version A is synthesized with the loop fully pipelined, resulting in a compact, sequential hardware architecture with a multiplier and an adder that are reused in each clock cycle. Version B, in contrast, is synthesized with the loop fully unrolled. The HLS compiler expands the loop into a large, parallel-dataflow architecture. Although both versions are algorithmically equivalent, their resulting hardware micro-architectures are drastically different at RTL. Verifying their equivalence often poses a challenge for formal verification tools.

Existing tools, such as AIGFuzz~\cite{Zhang2021}, AIGen~\cite{Jacobs2021}, FuzzBtor2~\cite{xiao2023fuzzbtor2} and BanditFuzz~\cite{scott2020banditfuzz} are unaware of the high-level algorithmic structures that are essential in the verification complexity. They often generate problems that are trivial to solve and therefore not suitable to assess the performance of the model checkers. \looseness=-1


\section{The \OurMethod Framework}
\label{sec:method}

\begin{figure}[htbp]
    \centering
    \includegraphics[width=1.0\textwidth]{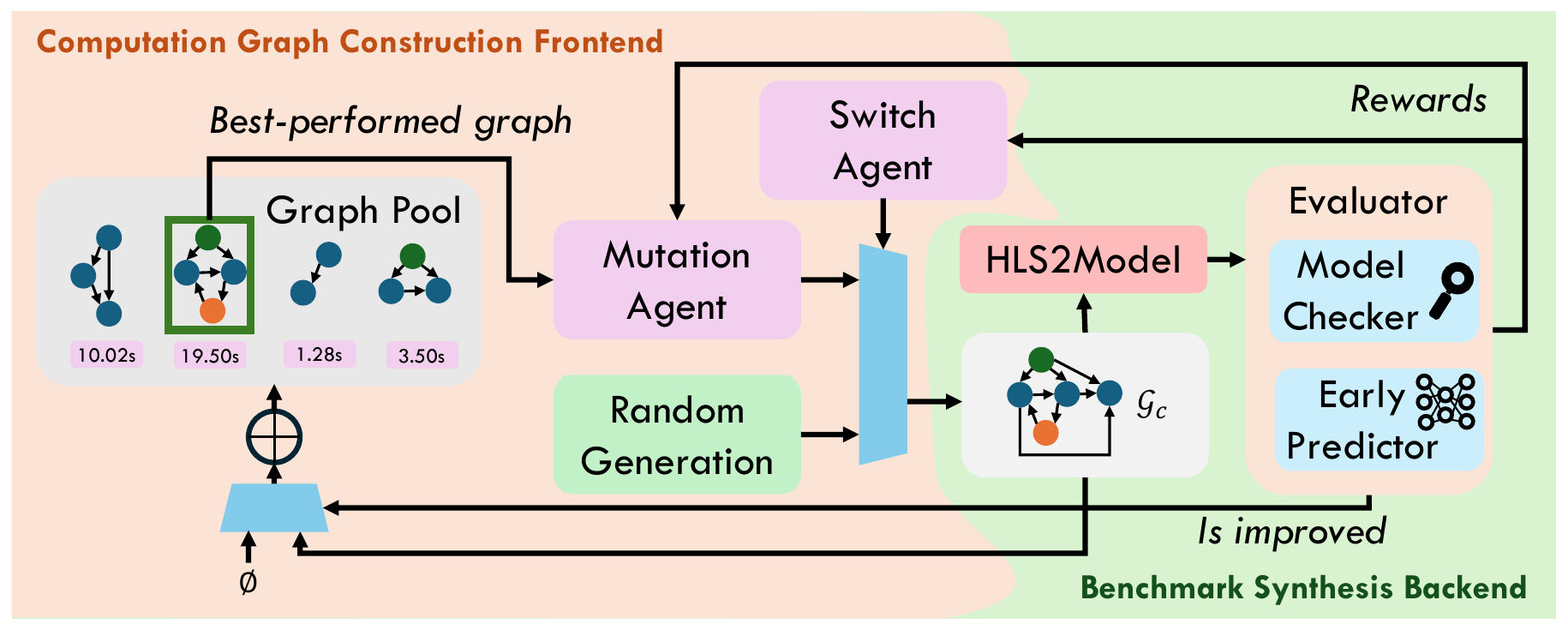}
    \caption{The overall workflow of our proposed \OurMethod framework.}
    \label{fig:overview}
\end{figure}

The overall workflow of \OurMethod is illustrated in Fig.~\ref{fig:overview}. The framework is composed of two main components: a \textbf{Computation Graph Construction Frontend} and a \textbf{Benchmark Synthesis Backend}.

The frontend is responsible for generating a computation graph, a data structure used to represent the HLS design under verification in our \OurMethod{} framework. It also controls the high-level structure of the HLS-generated hardware. The backend takes the computation graph as input, and processes the computation graph with an \HLSModel{} tool. This \HLSModel{} tool converts a computation graph into a corresponding model checking problem. The backend also contains an evaluator to evaluate the performance of the generated benchmark model checking problem, and generate feedback to the frontend of \OurMethod{}.

In the remainder of this section, we will first introduce in detail the definition of the computation graph and how a selected computation graph is converted to a model checking problem in Section~\ref{subsec:comp_graph_repr} and Section~\ref{subsec:bm_syn}, respectively. Then, in Section~\ref{subsec:rl-guided-generation}, we will introduce the detailed RL-guided process in the frontend to generate the computation graph under evaluation. In this Section, we will also introduce the technique to generate the reward signal for the frontend computation graph generation part. 


\subsection{Computation Graph Representation}\label{subsec:comp_graph_repr}
\begin{figure}[ht]
    \centering
    \includegraphics[width=\columnwidth]{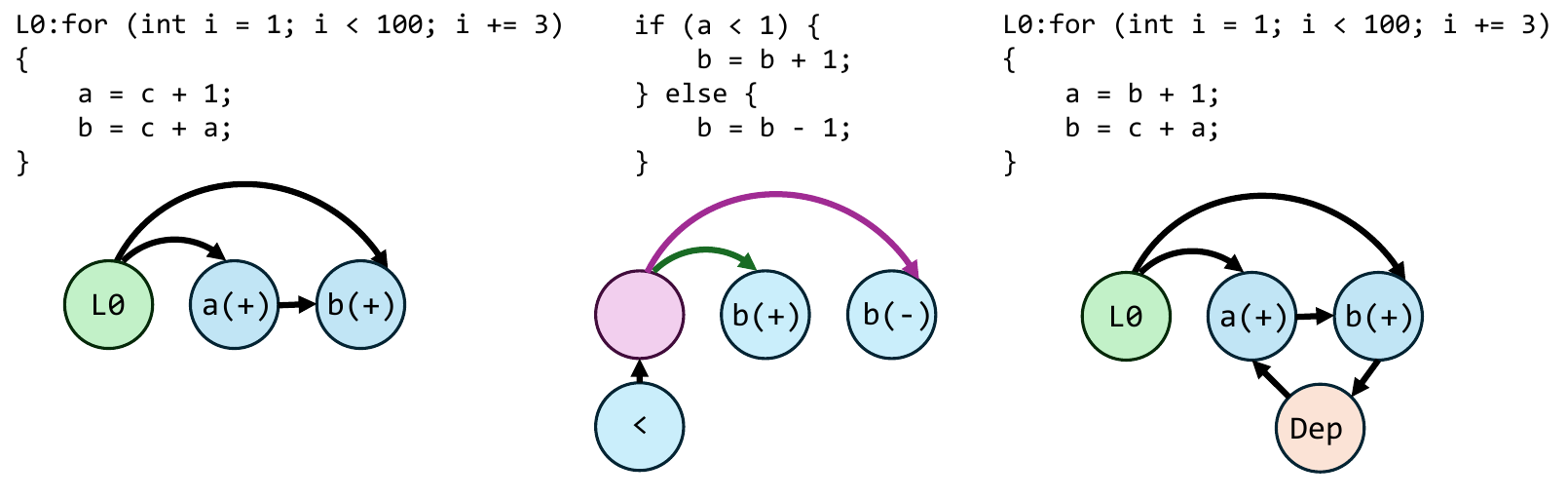}
    \caption{
    Examples of different type of nodes in a computation graph with the corresponding code snippet, where \texttt{c} is a constant.
    }
    \label{fig:action_space}
\end{figure}

At the core of our framework is a directed graph representation of a program, which we call a computation graph. The nodes represent operations or control structures, and the edges represent data flow or membership of code blocks. We define four types of nodes, with examples in Fig.~\ref{fig:action_space}.
During graph construction, the RL agent will attempt to add these types of nodes, representing corresponding graph transformation operations, which will be discussed in Section~\ref{subsec:rl-guided-generation}. The four types of nodes are:

\begin{itemize}
    \item \textbf{OpNode:} represents a computation, such as arithmetic (\texttt{ADD}, \texttt{MUL}), bitwise (\texttt{AND}, \texttt{XOR}), or comparison (\texttt{EQ}, \texttt{LT}) operations. Each OpNode has attributes defining its data type (e.g., \texttt{ap\_int<32>}, representing a 32-bit integer) and the operation type.
    \item \textbf{LoopNode:} represents a for-loop, with attributes for the start index, the ending index, and the step size. It also stores HLS pragma information, such as whether the loop is pipelined or unrolled.
    \item \textbf{BranchNode:} represents an if-then-else conditional branch. It is controlled by a predecessor OpNode whose result is treated as the branch condition.
    \item \textbf{DepNode:}  is a special node, representing an
    inter-iteration data dependency within a loop (e.g., $a[i] = a[i-1] + c$, where $i$ is the loop variable). There is also such an example in the last graph in Fig.~\ref{fig:action_space} where $a$ depends on the value of $b$ in the previous iteration. This is crucial for creating complex sequential behavior that is often challenging for model checkers. 
\end{itemize}

\definecolor{opnodebg}{RGB}{193,229,245}
\definecolor{loopnodebg}{RGB}{194,241,200}
\definecolor{branchnodebg}{RGB}{242,207,238}
\definecolor{depnodebg}{RGB}{251,227,214}
\begin{table}[t]
    \centering
    \caption{Node types and their attributes.}

    {\setlength{\aboverulesep}{0pt}%
     \setlength{\belowrulesep}{0pt}%
     \resizebox{\columnwidth}{!}{%
     \begin{tabular}{@{}l p{0.78\columnwidth}@{}}
        \toprule
        \textbf{Node type} & \textbf{Attributes} \\
        \midrule
        \cellcolor{opnodebg}\textbf{OpNode} & operation type, data width, data type, rounding mode, saturation mode, and the integer value bits count \\
        \cellcolor{loopnodebg}\textbf{LoopNode} & whether pipelined, whether flattened, loop variable information (including starting index, ending index, and the step size), unrolling mode (unroll factor, whether fully unrolled) \\
        \cellcolor{branchnodebg}\textbf{BranchNode} & the branch condition \\
        \cellcolor{depnodebg}\textbf{DepNode} & N/A \\
        \bottomrule
     \end{tabular}%
     }%
    }
    \label{tab:node_attri}
\end{table}

As presented above, a node carries the associated attributes listed in Table~\ref{tab:node_attri}.
These attributes summarize the customizations that can be done to an HLS-generated design. They encompass most of the design options for fixed-point datatypes and loops defined in the HLS tool, enabling the generated computation graphs to capture a rich variety of HLS design behaviors.
It should be noted that in our design of the computation graph, \textit{LoopNode} and \textit{BranchNode} are also used to represent the corresponding code block. The successor nodes of these nodes are the contents of the code block. 

\subsection{Benchmark Synthesis from Computation Graphs}\label{subsec:bm_syn}

A computation graph can be transformed into a model checking problem using the \HLSModel{} process. Once the computation graph is finalized, \HLSModel{} constructs two functionally equivalent hardware designs,  generated under distinct HLS optimization directives (pragams). These differing settings result in hardware implementations that are structurally divergent but  generate the identical computation result, enabling functional equivalence checking between them.

Note that the equivalence query in our generated problem is high-level and functional, representing a flexible notion of equivalence rather than the traditional combinational/sequential equivalence that requires either exactly the same combinational logic functions or cycle-by-cycle matching. As hardware verification tasks are mostly checking a certain kind of match between a specification and an implementation, which can often be formulated as a certain notion of equivalence, the equivalence checking problem we constructed mimic this flexible notion of equivalence and is therefore representative in hardware verification.

\begin{figure}[ht]
    \centering
    \includegraphics[width=1.0\columnwidth]{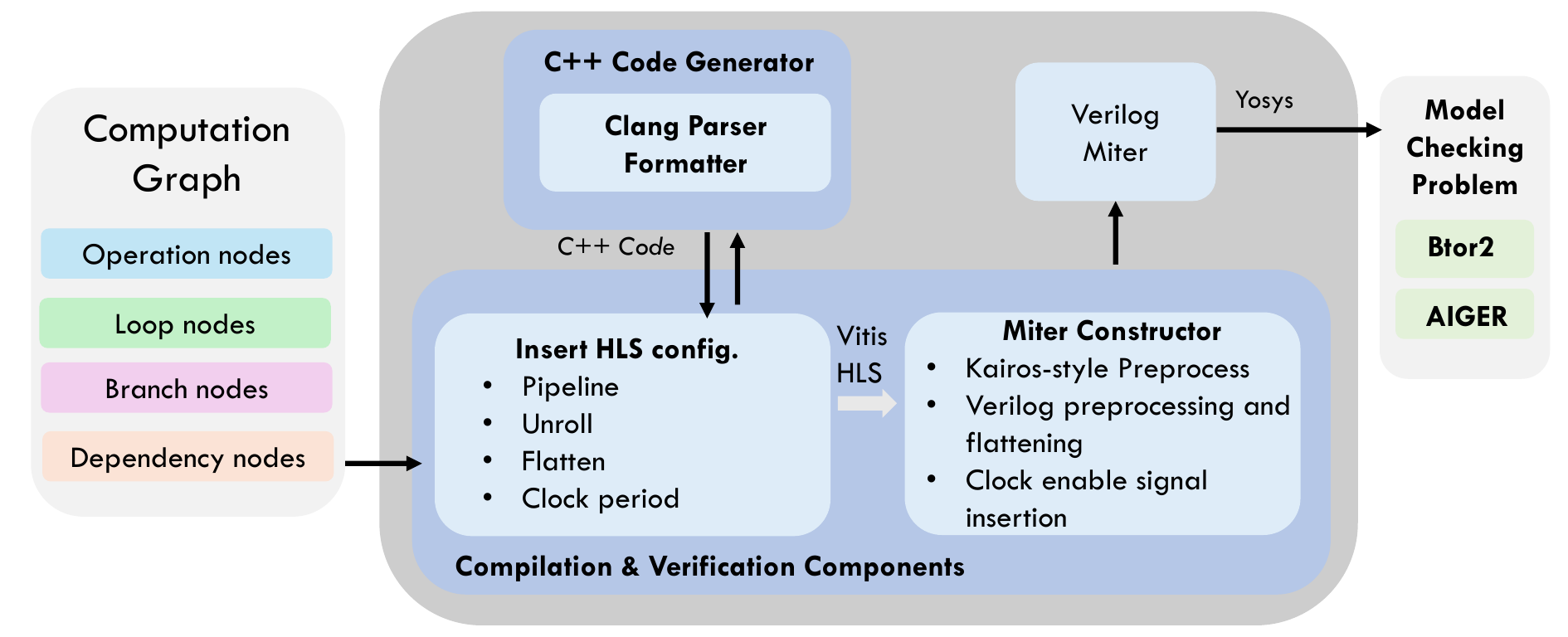}
    \caption{The backend workflow: synthesizing computation graphs to hardware designs.}
    \label{fig:rl_flow}
\end{figure}

As shown in Fig.~\ref{fig:rl_flow}, the \HLSModel{} can be split into four steps: C++ code generation, diversified HLS pragma insertion, miter generation, and model checking problem compilation.

\begin{enumerate}
    \item \textbf{C++ Code Generation:} The computation graph is traversed to generate a base C++ source file with the support of HLS-compatible arbitrary precision data types (e.g., \texttt{ap\_int} and \texttt{ap\_fixed}, representing the integer or fixed-point data types).

    \item \textbf{HLS Pragma Insertion:} From the input computation graph, we create two functionally equivalent but structurally distinct variants by applying different sets of HLS pragmas. This is the primary mechanism for creating structural divergence for meaningful and challenging model checking problems. Our framework employs two distinct optimization strategies to generate two versions of hardware implementations for the computation graph:
    \begin{itemize}
        \item \textbf{Basic:} This version is generated with minimal optimization pragmas. For instance, loops are synthesized using the HLS tool's default sequential scheduling without pipelining, and a relaxed target clock period is specified. 
        \item \textbf{Optimized:} This version is aggressively optimized. For nested loops, inner loops are explicitly pipelined to maximize throughput, while outer loops are flattened if the nested loop is a perfect nested loop, i.e., there is no instruction between the execution of the outer and inner loop. Moreover, a stringent target clock period is enforced, encouraging the HLS compiler to insert more registers to break possible critical paths in the design. This guides the HLS tool towards a more parallel, high-performance architecture.
    \end{itemize}
    The application of these divergent strategies to the same source graph results in two distinct C++ files, which serve as the inputs for the subsequent equivalence checking problem.

    \item \textbf{HLS Synthesis and Miter Generation:} Each C++ variant is synthesized into a Verilog module using a commercial HLS compiler~\cite{xilinx2021vivado}. 
    The synthesized designs are then instantiated within a top-level \textit{miter circuit} using the method in KAIROS~\cite{piccolboni2019kairos}, which is a functional equivalence checking framework that synchronizes the valid result production of HLS designs. The miter circuit connects their respective inputs to a common set of primary inputs and compares their outputs at the proper time. A dedicated single-bit output, \texttt{unsafe\_signal} is added, which is asserted if and only if a functional discrepancy between the two designs is detected.

    \item \textbf{Conversion to Verifier-Ready Format:} Finally, the Verilog miter circuit is processed through a logic synthesis flow~\cite{wolf2013yosys} that converts the circuit into formats that can be accepted by the model checker, namely AIGER and BTOR2 format. This yields the final benchmark artifact.
\end{enumerate}

\subsection{Frontend RL-Guided Graph Generation}
\label{subsec:rl-guided-generation}
To discover computation graphs that produce challenging benchmarks, we approach the generation process using the reinforcement learning (RL) method. Our goal is to build RL agents to navigate the vast search space of possible graph structures and identify those that, when synthesized, result in computationally challenging verification problems for a targeted model checker. The entire process is guided by a hybrid evolutionary strategy, detailed in Algorithm~\ref{alg:banditfuzz}, 
We will explain this method in two parts: (1) how different graph structures are constructed and (2) how rewards are computed to guide the RL agent to explore these different graph structures.

\begin{algorithm}[t]
\caption{The \OurMethod Algorithm}
\label{alg:banditfuzz}
\begin{algorithmic}[1]
\Require Maximum iterations $M_{max}$, Initial pool size $N_{init}$
\Ensure A pool $\mathcal{P}$ of hard-to-solve benchmarks (graphs)

\Statex
\Procedure{Evolve}{$M_{max}, N_{init}$}
    \State $\mathcal{P} \gets \Call{InitializePool}{N_{init}}$
    \State $\pi_{action}, \pi_{strategy} \gets \text{InitializeAgents()}$
    
    \While{$|\mathcal{P}| < M_{max}$}
        \State $S \gets \pi_{strategy}.\text{select\_action()}$ \Comment{Choose mutate or generate new graph}
        \If{$S = \text{Mutate}$ \textbf{and} $\mathcal{P} \neq \emptyset$}
            \State $(\mathcal{G}_{p}, R_{p}) \gets \text{SelectFromPool}(\mathcal{P})$
            \State $a \gets \pi_{action}.\text{select\_action()}$
            \State $\mathcal{G}_{c} \gets \text{Mutate}(\mathcal{G}_{p}, a)$
            \State $R_{new}, \text{status} \gets \text{Evaluate}(\mathcal{G}_{c})$ 
            \State $R_{base} \gets R_{p}$ 
        \Else \Comment{$S = \text{Generate}$}
            \State $\mathcal{G}_{c} \gets \text{GenerateFreshGraph()}$
            \State $R_{new}, \text{status} \gets \text{Evaluate}(\mathcal{G}_{c})$
            \State $R_{base} \gets \text{AveragePoolPerformance}(\mathcal{P})$
        \EndIf
        
        \If{status is \textbf{success}}
            \State \text{improved} $\gets R_{new} > R_{base}$
            \State $\pi_{strategy}.\text{reward}(\text{improved})$
            \If{$S = \text{Mutate}$} $\pi_{action}.\text{reward}(\text{improved})$ \EndIf
            \If{improved} $\mathcal{P} \gets \mathcal{P} \cup \{(\mathcal{G}_{c}, R_{new}, \dots)\}$ \EndIf
        \EndIf
    \EndWhile
\EndProcedure
\end{algorithmic}
\end{algorithm}

\minisection{Multi-Agent Strategy for Graph Exploration.}
We adopt a multi-agent reinforcement learning (MARL) approach inspired by BanditFuzz~\cite{scott2020banditfuzz}, framing the generation process as a Multi-Armed Bandit (MAB) problem. There are in total two distinct agents that collaborate to balance exploration and exploitation: a high-level \textit{Switch Agent} and a low-level \textit{Mutation Agent}.
This arrangement efficiently decomposes the learning task, allowing the Mutation Agent to focus purely on learning which structural operators yield hardness, while the Switch Agent independently manages population diversity, leading to faster convergence than a monolithic single-agent formulation and helps better escape local optima.  Both agents employ the  Thompson Sampling method~\cite{agrawal2012analysis}.
%
%
The overall workflow, also depicted in Fig.~\ref{fig:overview}, maintains a \textit{graph pool} ($\mathcal{P}$) of promising graphs and iteratively seeks to improve it.

The algorithm takes two parameters as input: the maximum number of iterations and the initial pool size.
The main loop (Line 4) is responsible for graph generation and it iterates until the pool reaches a target size. At each step, the \textit{Switch Agent} first selects one of two strategies (Line 5):
\begin{itemize}
    \item \textbf{Mutate:} This exploitation-focused strategy can be chosen if the pool is not empty (Line 6). It selects a promising graph $(\mathcal{G}_{p})$ from the pool, prioritizing the graph with the highest predicted solving time (Line 7). The core mutation process is modeled as a Markov Decision Process (MDP), where the current graph is the state $s_t = \mathcal{G}_{p}$ of this MDP. The \textit{Mutation Agent} then selects an action $\vec{a_t}$ (Line 8), which is a vector of graph transformation operators (e.g., adding a node from \texttt{<OpNode, BranchNode, LoopNode, DepNode>}) to apply to $\mathcal{G}_{p}$, yielding a new graph $\mathcal{G}_{c}$ (Line 9).
    A pair of rewards are computed (Line 10-11), where the new reward is evaluated following the method \texttt{Evaluate}, which will be discussed later, and the baseline reward for this action is the known parent's reward, $R_p$. These two will be compared to help improve the policy of the agent.
    \item \textbf{Generate:} This exploration-focused strategy creates a completely new, randomly constructed graph from scratch (Line 13) using \texttt{GenerateFreshGraph}. This function takes a hyperparameter, which is the sequence length of random construction actions. This hyperparameter is computed dynamically based on the average size of the current high-performing pool (capped at 40 actions) to strictly enforce the ``small-but-hard'' criterion.
    We will also evaluate the generated graph using the \texttt{Evaluate} function, which forms the new reward (Line 14),
    while the baseline reward for this choice is the average performance of the entire pool (Line 15), providing a general measure of quality.
\end{itemize}

After a new graph $\mathcal{G}_{c}$ is constructed either by Mutate or Generate, a reward $R_{new}$ is determined, which is associated to the predicted solving time of the synthesized benchmark. If this reward is higher than the calculated baseline ($R_{new} > R_{base}$), the action is considered an ``improvement'' (Line 18). A positive feedback signal is then sent to update the agents' policies (Lines 19-20), and the improved graph is added to the pool (Line 22). This reward-driven loop allows the agents to learn which strategies and mutations have historically yielded difficult benchmarks.

\minisection{Early Prediction of Benchmark Difficulty.}
A key challenge in our RL loop is the high cost of obtaining an accurate reward signal, as a full model checking run is too slow for rapid training iterations. To address this, we developed a lightweight model to predict a benchmark's solving time based on features extracted from a brief fixed-depth formal analysis.

This predictor is a gradient boosting regression model (XGBoost)~\cite{Chen2016xgboost} trained offline on a curated dataset. We collected data from the HWMCC'20 \& '24 competitions and constructed our dataset by selecting only those cases that could be solved within the first five recursion frames of PDR and completed in under 120 seconds. This selection criterion ensures the dataset is populated with benchmarks that yield informative short-run dynamic features relevant to our prediction task. The model learns a mapping from a vector of static and dynamic circuit features (see Table~\ref{tab:predictor-features}) to the final solving time. On our test set, the predictor achieved a high degree of accuracy, with $R^2$ scores of 0.60 for rIC3, 0.58 for IC3ref, and 0.46 for ABC PDR. This validates its use as a reliable reward proxy, enabling efficient RL-guided synthesis.

\begin{table}[h]
    \centering
    \caption{Key features used by the model checking difficulty predictor model.}
    \resizebox{\columnwidth}{!}{%
    \begin{tabular}{@{}lll@{}}
        \toprule
        \textbf{Category} & \textbf{Feature Class} & \textbf{Description and Rationale} \\
        \midrule
        \multirow{4}{*}{\textbf{Static}}
            & Size Metrics & Basic indicators of circuit scale (PI, flip-flop, AIG node counts). \\
            & Structural Complexity & Measures of functional units (MUX, XOR, adder counts). \\
            & Topological Depth & Captures the longest path and overall circuit topology (max/avg depth). \\
            & Sequential Connectivity & Quantifies dependencies between state elements (flop fanout statistics). \\
        \midrule
        \multirow{3}{*}{\textbf{Dynamic}}
            & PDR Proof Complexity & Internal statistics from a short PDR run (e.g., number of generated \\
            &                      & clauses, proof obligations, and SAT solver calls) to gauge initial proof effort. \\
            \cmidrule(l){2-3}
            & PDR Runtime Profile & Time distribution across PDR sub-procedures (e.g., total SAT solving, \\
            &                     & clause generalization, and clause pushing) to identify computational bottlenecks. \\
            \cmidrule(l){2-3}
            & Clause Generation Statistics & Metrics summarizing the distribution of new clauses per frame \\
            &                              & (e.g., mean, standard deviation, max), reflecting inductive difficulty. \\
        \bottomrule
    \end{tabular}%
    }
    \label{tab:predictor-features}
\end{table}

The features consist of static features and the dynamic ones.
\textbf{Static features} are derived directly from the AIG circuit's structural properties without executing any verification tasks, providing a quick snapshot of its size and complexity. \textbf{Dynamic features}, conversely, are designed to capture the circuit's effect on the formal proof search process, thereby revealing deeper properties related to its inductive complexity. To acquire these dynamic features efficiently, we perform a constrained fixed-depth (e.g., for 5 frames) run of the PDR algorithm. This short execution, while not intended to complete the proof, yields a rich set of predictive indicators. As detailed in Table~\ref{tab:predictor-features}, the dynamic features are grouped into three classes. As the PDR runtime profile is easy to understand, here we  only explain the proof complexity and generation statistics in detail:
\begin{itemize}
    \item \textbf{PDR Proof Complexity:} These features serve as direct measures of the proof work done by the PDR solver in its initial exploration. A high number of generated clauses suggests that the state space is complex and requires many lemmas to block bad states. Similarly, a large volume of proof obligations and SAT solver calls indicates that the solver is struggling to generalize lemmas and push them to later frames, a hallmark of a difficult inductive proof.

    \item \textbf{Clause Generation Statistics:} These features capture the core of inductive reasoning in PDR. Instead of merely counting new clauses, we analyze their \textit{behavior} across frames. A key operation in PDR is ``pushing'' a learned clause (a lemma) to the highest possible frame, signifying its general inductive strength. When many new clauses are generated at a particular frame, it often indicates that existing lemmas are too specific and cannot be pushed, forcing the solver to re-discover invariants. We compute robust statistical metrics over the sequence of newly generated clauses per frame. A high mean or a large variance in this sequence signals significant inductive difficulty, as it implies the solver is struggling to find generalizable lemmas and is instead repeatedly generating new, localized ones that fail to propagate. This provides a direct measure of the proof's inductive complexity.
\end{itemize}

The predicted solving time from this model serves as the reward signal $R_{new}$ for our RL agents. This approach enables rapid iteration, allowing the agents to efficiently explore the vast search space of possible computation graphs without incurring the high cost of full model checking at each step.

%
%
\section{Experimental Evaluation}\label{sec:experiment}


Our experimental evaluation is designed to answer three research questions (Section~\ref{sec:RQ1}--\ref{sec:RQ3}) that assess the effectiveness of \OurMethod{} and the quality of the generated benchmarks.
\begin{itemize}
    \item \textbf{RQ1:} Does \OurMethod generate challenging benchmarks efficiently?
    \item \textbf{RQ2:} Can \OurMethod generate high-quality benchmarks?
    \item \textbf{RQ3:} Can \OurMethod distinguish performance differences between two SOTA model checkers?
\end{itemize}

\subsection{Experimental Setup}\label{sec:exp_setting}

\minisection{Target Model Checkers} 
We use four state-of-the-art hardware model checkers: \textbf{rIC3}~\cite{su2025ric3}, \textbf{ABC-PDR}~\cite{mishchenko2007abc}, \textbf{IC3Ref}~\cite{bradley2011sat} and \textbf{Pono}~\cite{mann2021pono} as the model checkers guiding the benchmark generation. The predicted solving time of these model checkers on the generated model checking problem serves as the reward signal for our RL agent. We evaluate the generated benchmarks with these model checkers as well. 

\minisection{Baselines.}
We compare \OurMethod{} against several main baseline tools that can automatically produce model checking problems:
\begin{itemize}
    \item \textbf{FuzzBtor}~\cite{xiao2023fuzzbtor2}: A state-of-the-art btor fuzzer that generate btor2 format problem for fuzzing model checker and SMT solvers.
    \item  \textbf{AIGen}~\cite{Jacobs2021}: A tool that randomly generates boolean functions together with latches using uniform random sampling over the space of all boolean functions.
    \item  \textbf{AIGFuzz}~\cite{Zhang2021}: A fuzzer that randomly generates AIGs.
\end{itemize}

\minisection{Environment.}
All experiments were conducted on a server equipped with an Intel~11th~Gen~CPU~i5-11400~@~4.400GHz and 128~GB of RAM, running Ubuntu~20.04. The timeout for each model checking run is set to 3600 seconds.


\subsection{RQ1: Does \OurMethod generate challenging benchmarks efficiently?}\label{sec:RQ1}

\begin{figure}[h]
    \centering
    \begin{subfigure}[b]{0.45\linewidth}
        \centering
        \includegraphics[width=\linewidth]{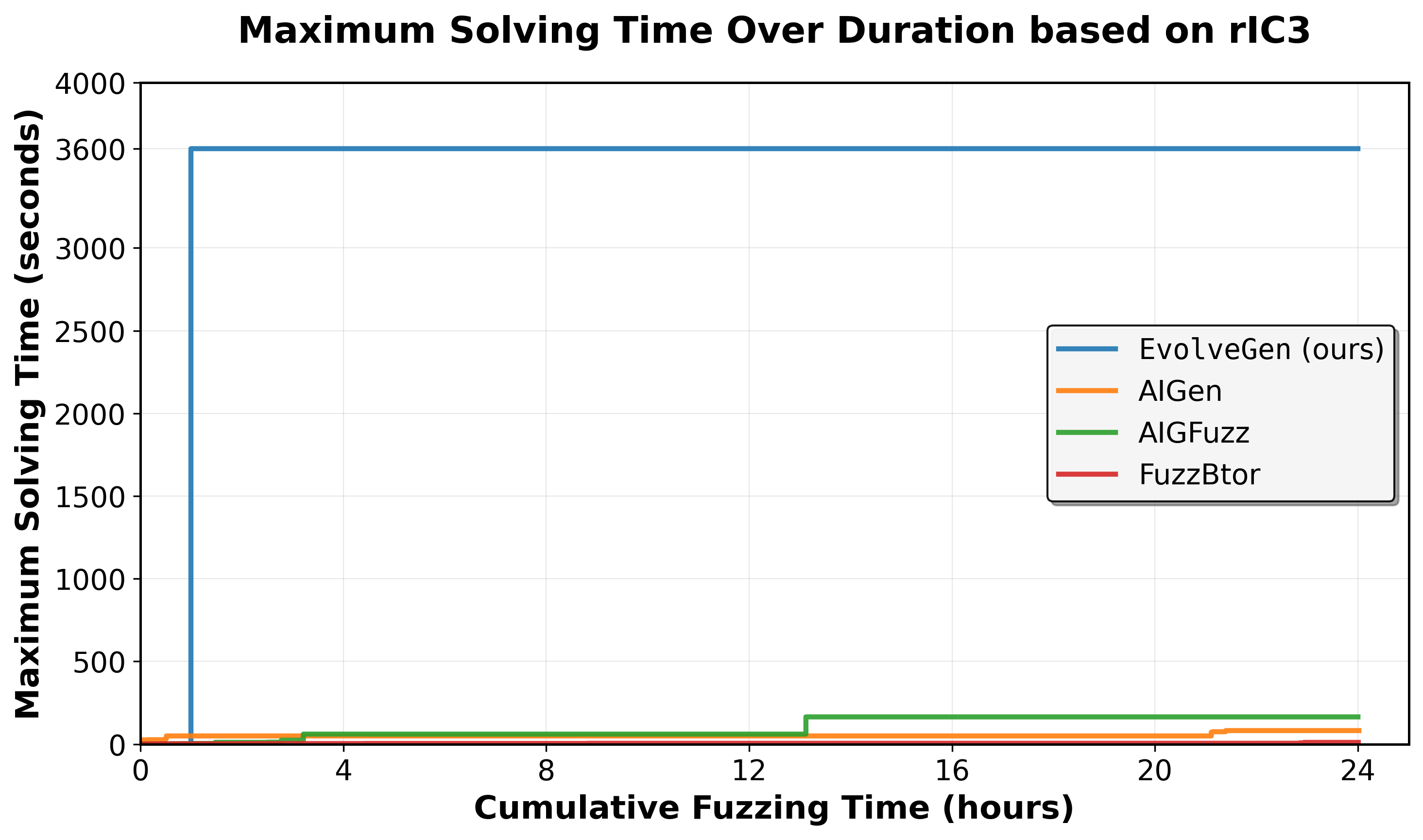}
        \caption{rIC3}
        \label{fig:rq1_ric3}
    \end{subfigure}
    \begin{subfigure}[b]{0.45\linewidth}
        \centering
        \includegraphics[width=\linewidth]{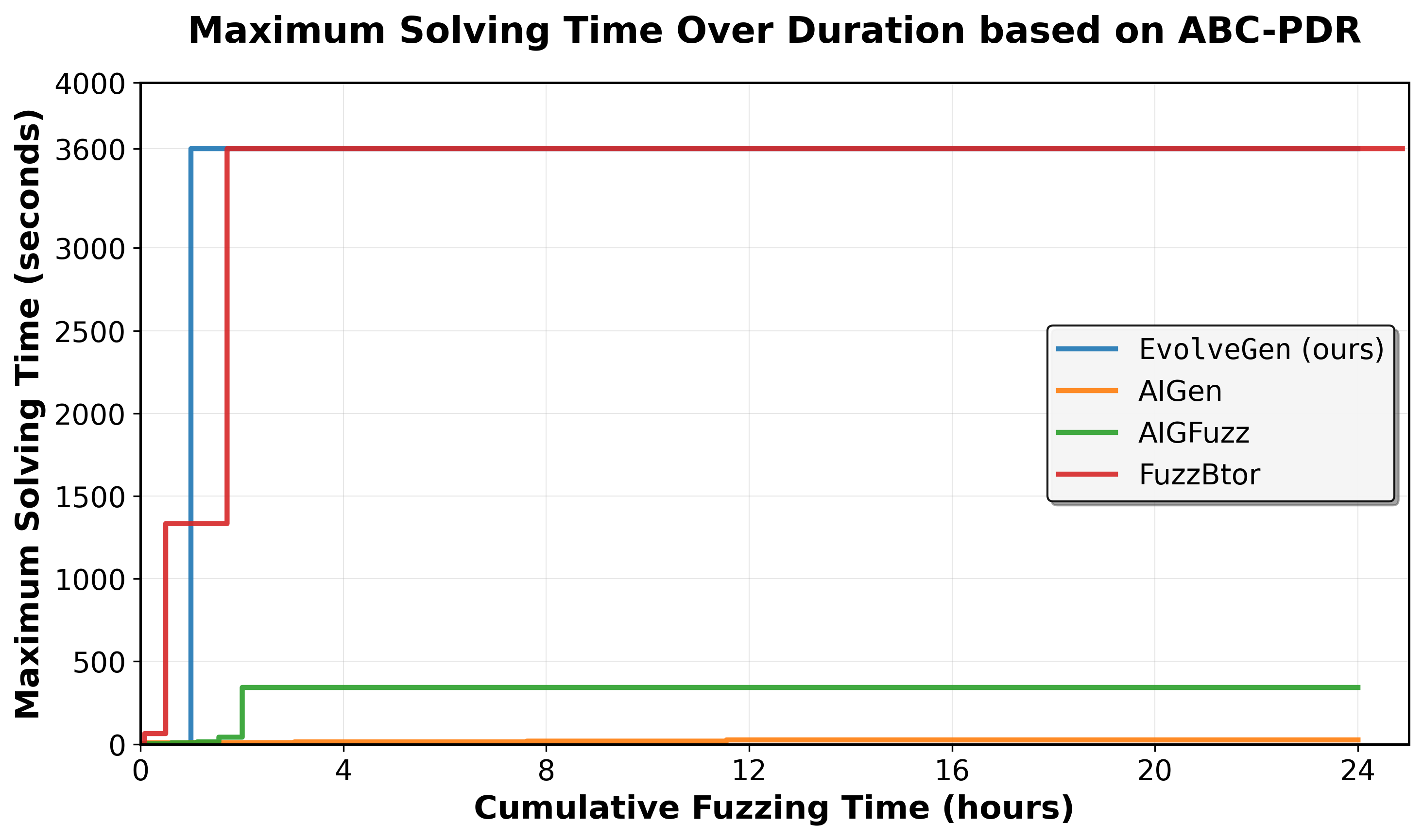}
        \caption{ABC-PDR}
        \label{fig:rq1_abc}
    \end{subfigure}
    \begin{subfigure}[b]{0.45\linewidth}
        \centering
        \includegraphics[width=\linewidth]{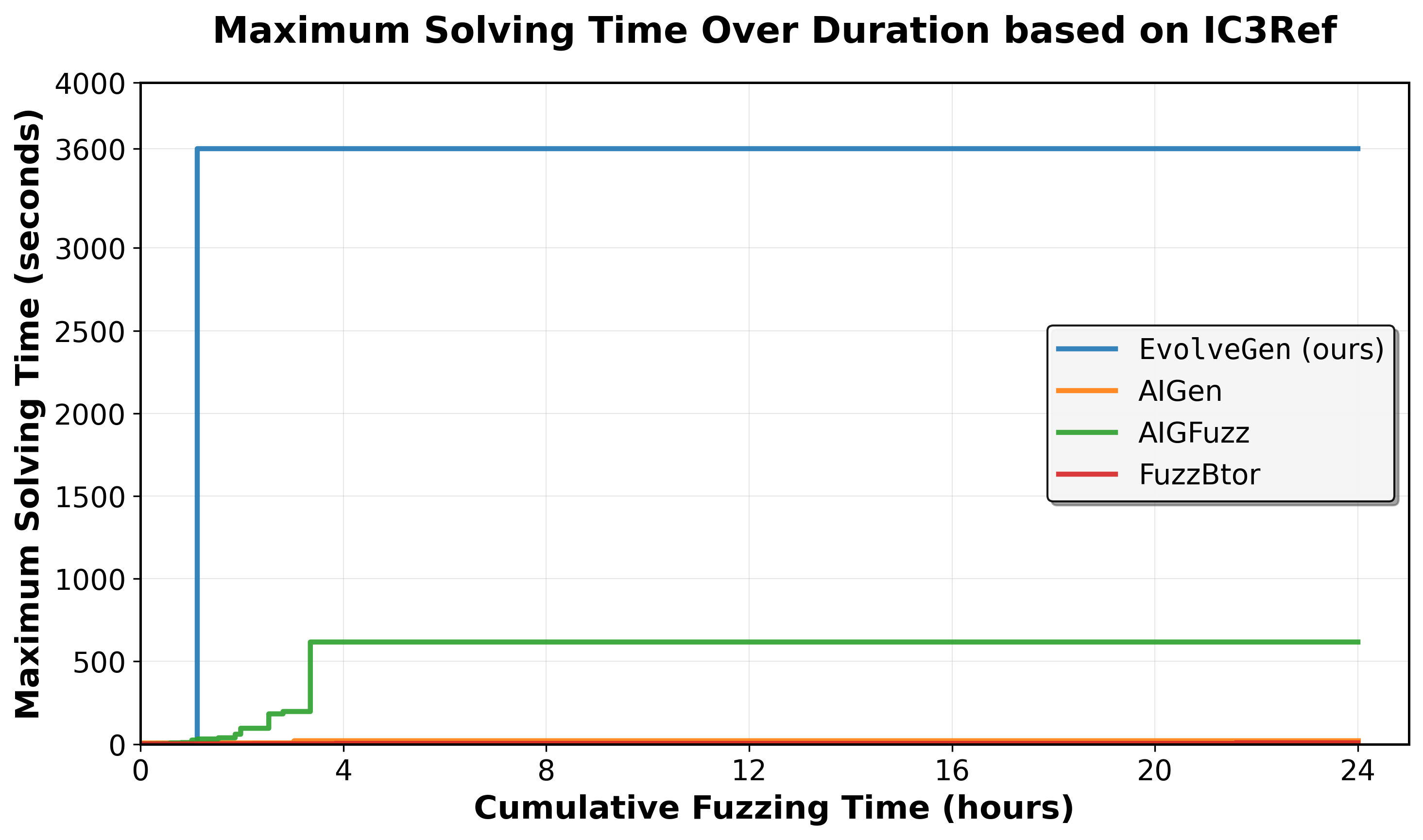}
        \caption{IC3Ref}
        \label{fig:rq1_ic3ref}
    \end{subfigure}
    \begin{subfigure}[b]{0.45\linewidth}
        \centering
        \includegraphics[width=\linewidth]{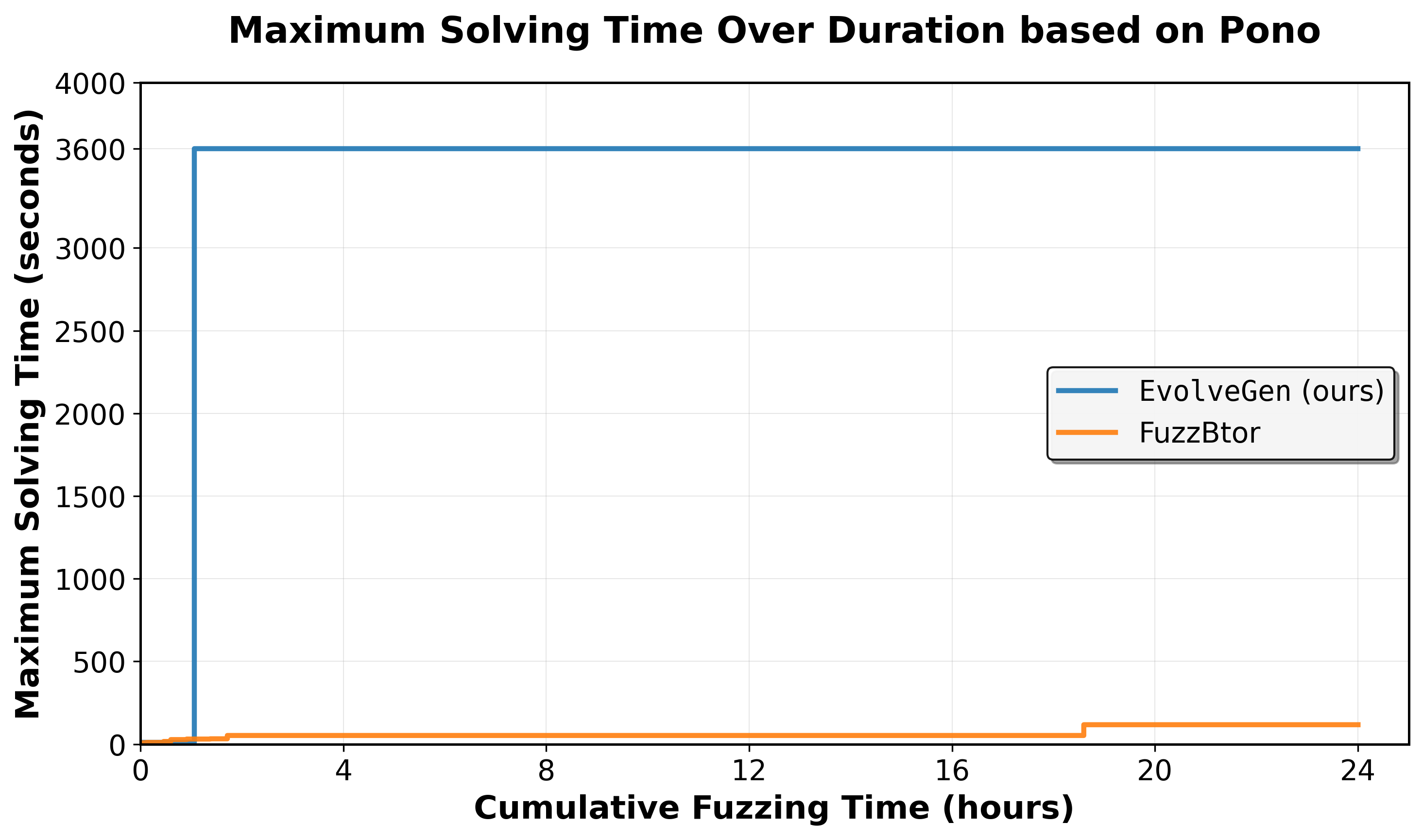}
        \caption{Pono}
        \label{fig:rq1_pono}
    \end{subfigure}
    \caption{Maximum solving time of derived model checking problems over time}
    \label{fig:rq1}
\end{figure}

The first research question discusses whether \OurMethod can quickly derive challenging benchmarks compared with existing methods. We run a 24-hour generation for each tool to find challenging cases. Fig.~\ref{fig:rq1} illustrates the evolution of the maximum solving time of generated model checking benchmarks over time. 
During the 24-hour generation process, we find that the baseline tools struggle to produce model checking problems that exceed a 10-minute solving time. In contrast, \OurMethod, guided by a RL-based generation strategy, can quickly identify patterns associated with hard-to-solve cases for a specific model checker by leveraging the reward signal indicating the expected solving time. This enables it to generate difficult model checking problems early in the generation process. In Fig.~\ref{fig:rq1_abc}, although FuzzBtor is eventually able to generate a benchmark that takes more than 1 hour to solve on the ABC-PDR model checker, it requires more time to discover such a challenging case compared to \OurMethod.

\subsection{RQ2: Can \OurMethod generate high-quality benchmarks?}\label{sec:RQ2}
High-quality benchmarks are defined as ``small-but-hard'', referring to circuits with modest sizes that nonetheless require substantial solving time, as discussed in Section~\ref{sec:motivation}. The heatmap in Fig.~\ref{fig:motivation-e}, which plots solving time against circuit size reveals that the existing benchmark suite lacks ``small-but-hard'' (high-quality) cases. To assess benchmark quality from \OurMethod, we use the quality ratio ($QR$), defined in Eq.~\ref{eq:quality_rate}. Table~\ref{tab:fuzzer_comparison} compares the $QR$ values of existing benchmark generation tools and \OurMethod{}. It reports the average circuit size and solving time of the 10 most difficult cases produced during the 24-hour period.

\begin{table}[htbp]
    \centering
    \caption{Comparison of different tools across various model checkers (for each model checker and each solver, we perform 24-hour generation and calculate the average of 10 most challenging cases, and $\overline{QR}$ is calculated by the ratio of the average solving time and the average benchmark size, and \textbf{normalized}). \textbf{Timeouts are calculated as 3600s}.}
    \begin{tabular}{l|l|c|c|c|c}
        \hline
        \textbf{Solver} & \textbf{Metric} & \textbf{AIGen} & \textbf{AIGFuzz} & \textbf{FuzzBtor} & \textbf{\OurMethod{}} \\
        \hline
        \multirow{3}{*}{ABC-PDR}
            & Size & 720179 & 26851 & 162548 & \cellcolor{lightblue}15279 \\
            & Time & 17.03 & 243.38 & \cellcolor{lightblue}3600.00& \cellcolor{lightblue}3600.00\\
            & $\overline{QR}$ & 1.00$\times10^{-4}$ & 3.84$\times10^{-2}$ & 9.36$\times10^{-2}$ & \cellcolor{lightblue}1.00\\
        \hline
        \multirow{3}{*}{IC3Ref}
            & Size & 719888 & 29079 & 194439 & \cellcolor{lightblue}16329\\
            & Time & 17.78 & 294.58 & 3.73 & \cellcolor{lightblue}3600.00\\
            & $\overline{QR}$ & 1.12$\times10^{-4}$ & 4.59$\times10^{-2}$ & 8.73$\times10^{-5}$ & \cellcolor{lightblue}1.00\\
        \hline
        \multirow{3}{*}{rIC3}
            & Size & 719930 & 29633 & 237740 & \cellcolor{lightblue}20217\\
            & Time & 69.75 & 58.99 & 5.95 & \cellcolor{lightblue}3600.00\\
            & $\overline{QR}$ & 5.44$\times10^{-4}$ & 1.12$\times10^{-2}$ & 1.40$\times10^{-4}$ & \cellcolor{lightblue}1.00\\
        \hline
        \multirow{3}{*}{Pono}
            & Size & -- & -- & 224363 & \cellcolor{lightblue}4052\\
            & Time & -- & -- & 64.41 & \cellcolor{lightblue}3465.77\\
            & $\overline{QR}$ & -- & -- & 3.36$\times10^{-4}$ & \cellcolor{lightblue}1.00\\
        \hline
    \end{tabular}
    
    \label{tab:fuzzer_comparison}
\end{table}

From the results, we observe that AIGen, although capable of generating large AIGs, produces benchmarks that are relatively easy to solve. Across all model checkers, the hardest cases require less than two minutes to be solved on average. AIGFuzz shows similar behavior: it tends to generate large circuits, but the resulting model checking problems are generally not challenging. FuzzBtor exhibits this pattern only on IC3Ref, rIC3, and Pono. Although FuzzBtor can generate benchmarks that stress the ABC-PDR solver (shown in RQ1, Fig.~\ref{fig:rq1_abc}), the difficulty primarily arises from the excessive circuit size rather than intrinsic complexity of state space.

In contrast, our proposed method, \OurMethod{}, consistently produces model checking problems that are significantly harder to solve, while maintaining a relatively compact circuit size. This demonstrates its effectiveness in generating high-quality benchmarks that better satisfy the ``small-but-hard'' criterion. Specifically, we define the quality of a generated instance as:

\begin{equation}\label{eq:quality_rate}
    QR = \frac{\text{Solving time of generated case}}{\text{AND gate count} + \text{Latch count}}
\end{equation}

\begin{figure}[h]
    \centering
    \begin{subfigure}[b]{0.23\textwidth}
        \includegraphics[width=\textwidth]{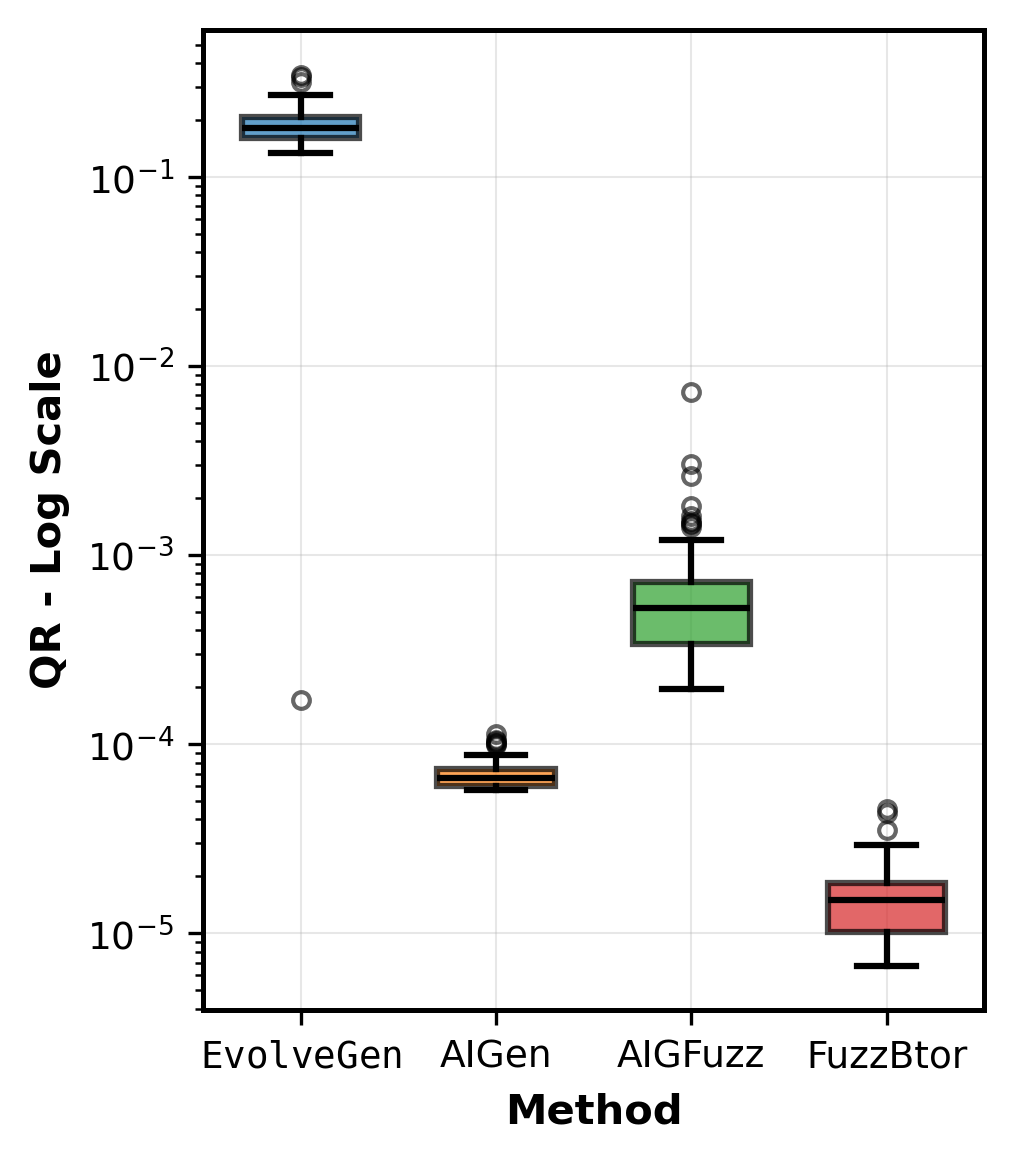}
        \caption{rIC3}
        \label{fig:rq2_ric3}
    \end{subfigure}
    \hfill
    \begin{subfigure}[b]{0.23\textwidth}
        \includegraphics[width=\textwidth]{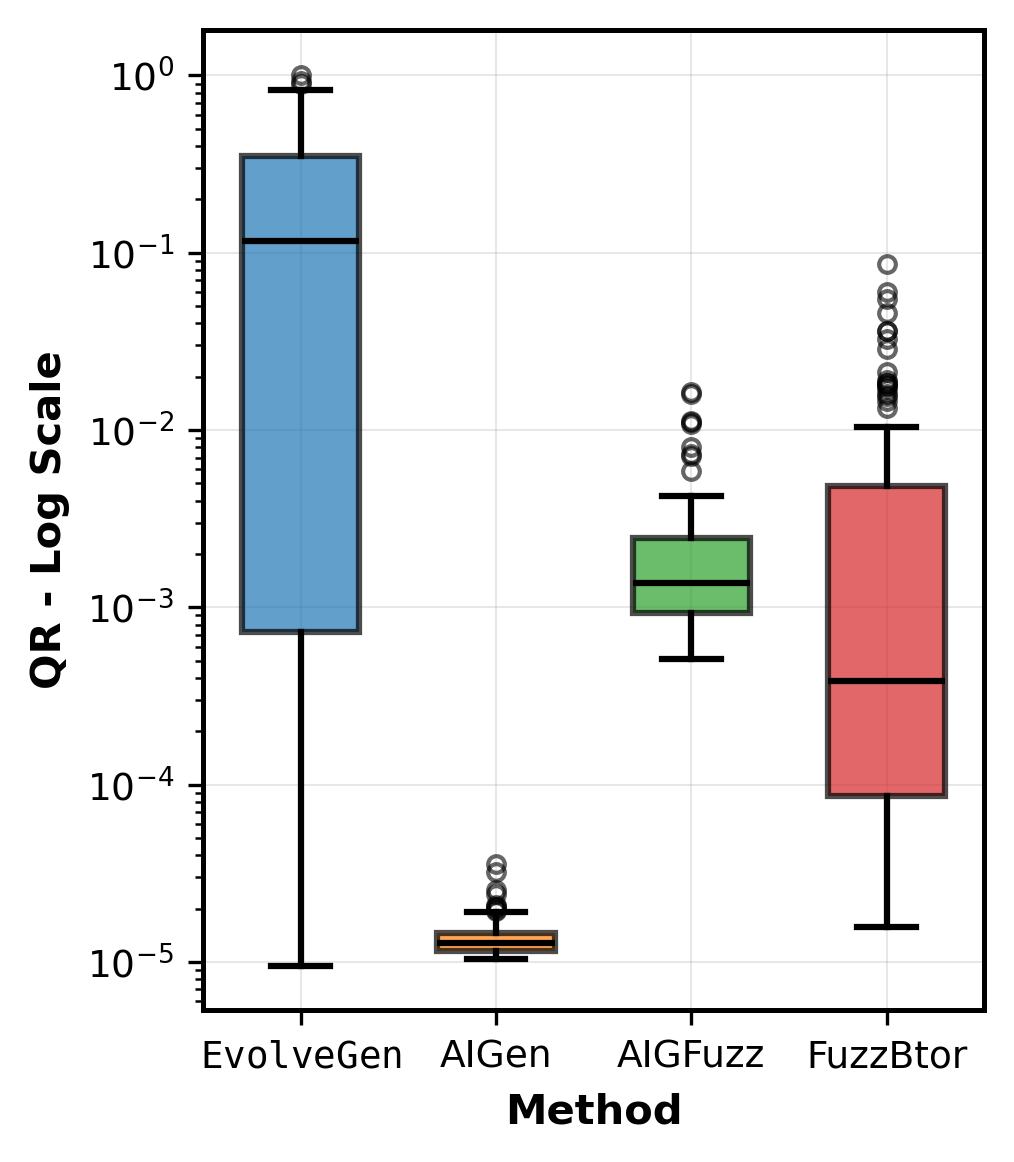}
        \caption{ABC-PDR}
        \label{fig:rq2_abc_pdr}
    \end{subfigure}
    \hfill
    \begin{subfigure}[b]{0.23\textwidth}
        \includegraphics[width=\textwidth]{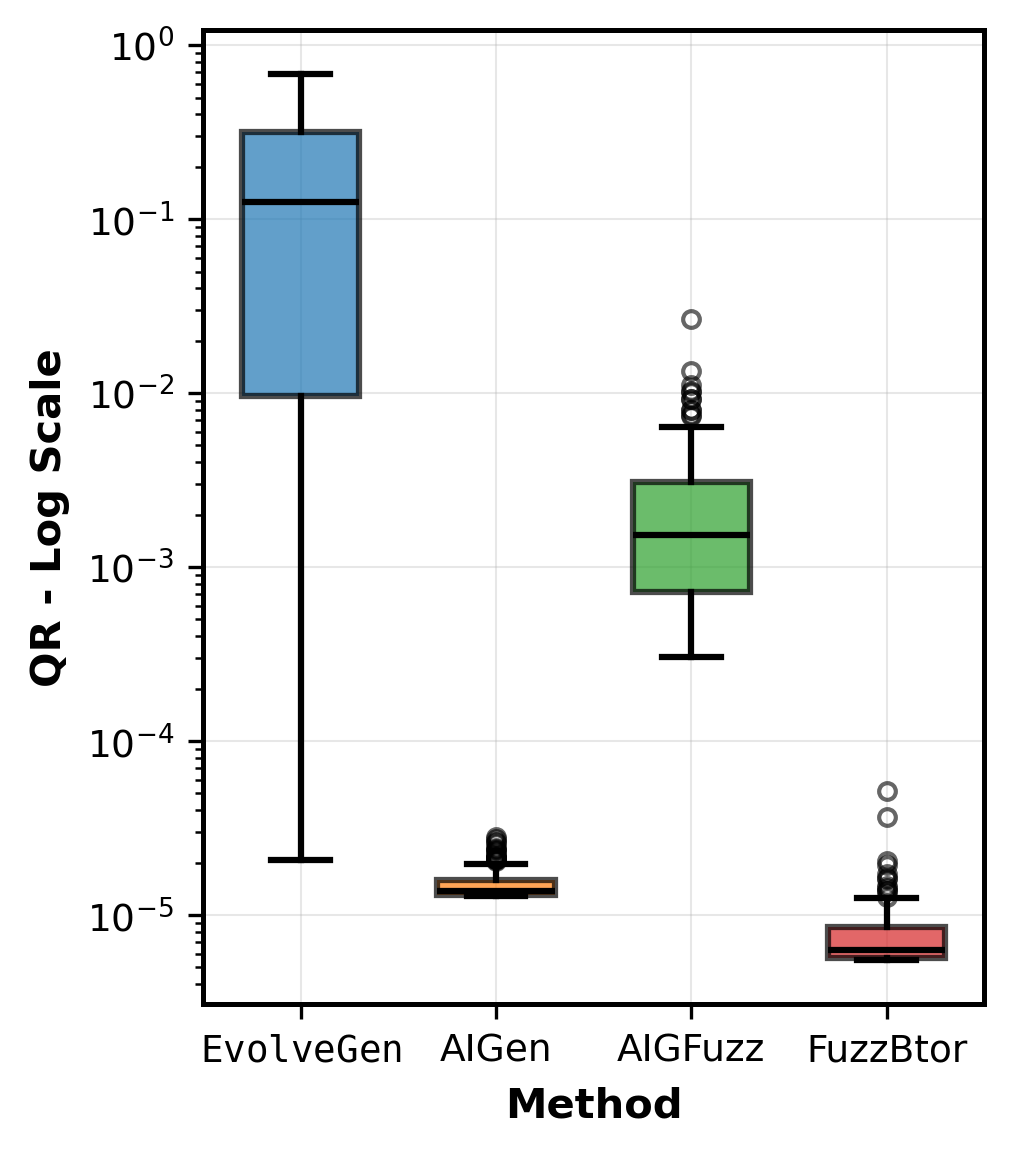}
        \caption{IC3Ref}
        \label{fig:rq2_ic3ref}
    \end{subfigure}
    \hfill
    \begin{subfigure}[b]{0.23\textwidth}
        \includegraphics[width=\textwidth]{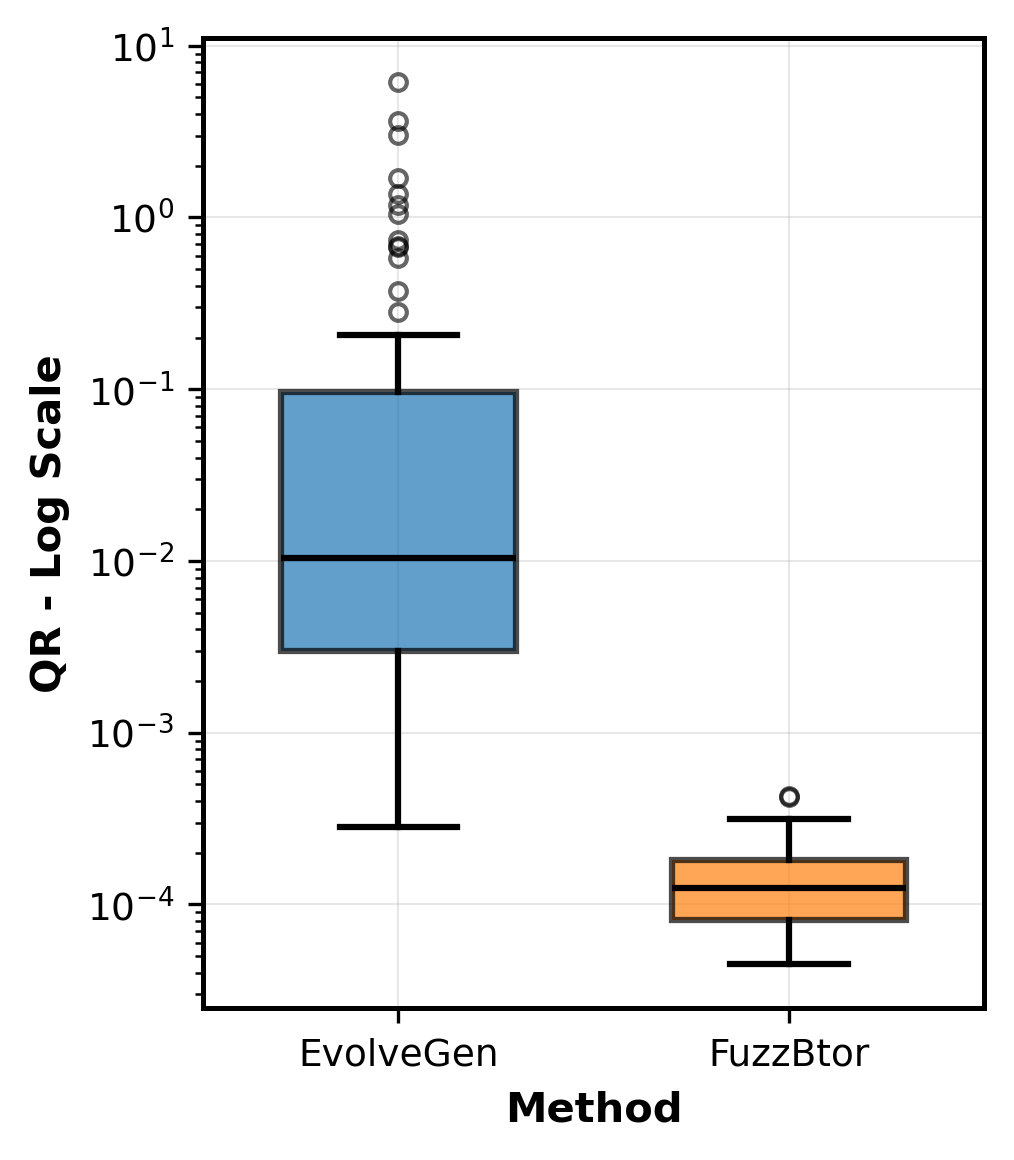}
        \caption{Pono}
        \label{fig:rq2_pono}
    \end{subfigure}
    \caption{Distribution of $QR$, for baseline tools, i.e., AIGen, AIGFuzz, and FuzzBtor, we pick the top 100 slowest generated benchmarks, and for \OurMethod, we use the whole generated benchmark suite.}
    \label{fig:rq2}
\end{figure}

Fig.~\ref{fig:rq2} provides the distribution of $QR$ for the top 100 most challenging cases generated by baseline tools and all generated cases in our model checkers. \OurMethod outperforms baseline tools by at least an order of magnitude in QR in most of the cases, indicating that our method can generate small but challenging model checking problems. Although there are a few simple cases from \OurMethod that do not have high quality, it is primarily due to the misprediction of the early predictor. Nevertheless, owing to the short solving time, such cases can be efficiently identified and filtered out through quick testing, and therefore have minimal impact on the overall quality of the generated benchmark suite.\looseness=-1

\subsection{RQ3: Can \OurMethod distinguish performance differences between two SOTA model checkers?}\label{sec:RQ3}
\begin{figure}
    \centering
    \includegraphics[width=0.5\linewidth]{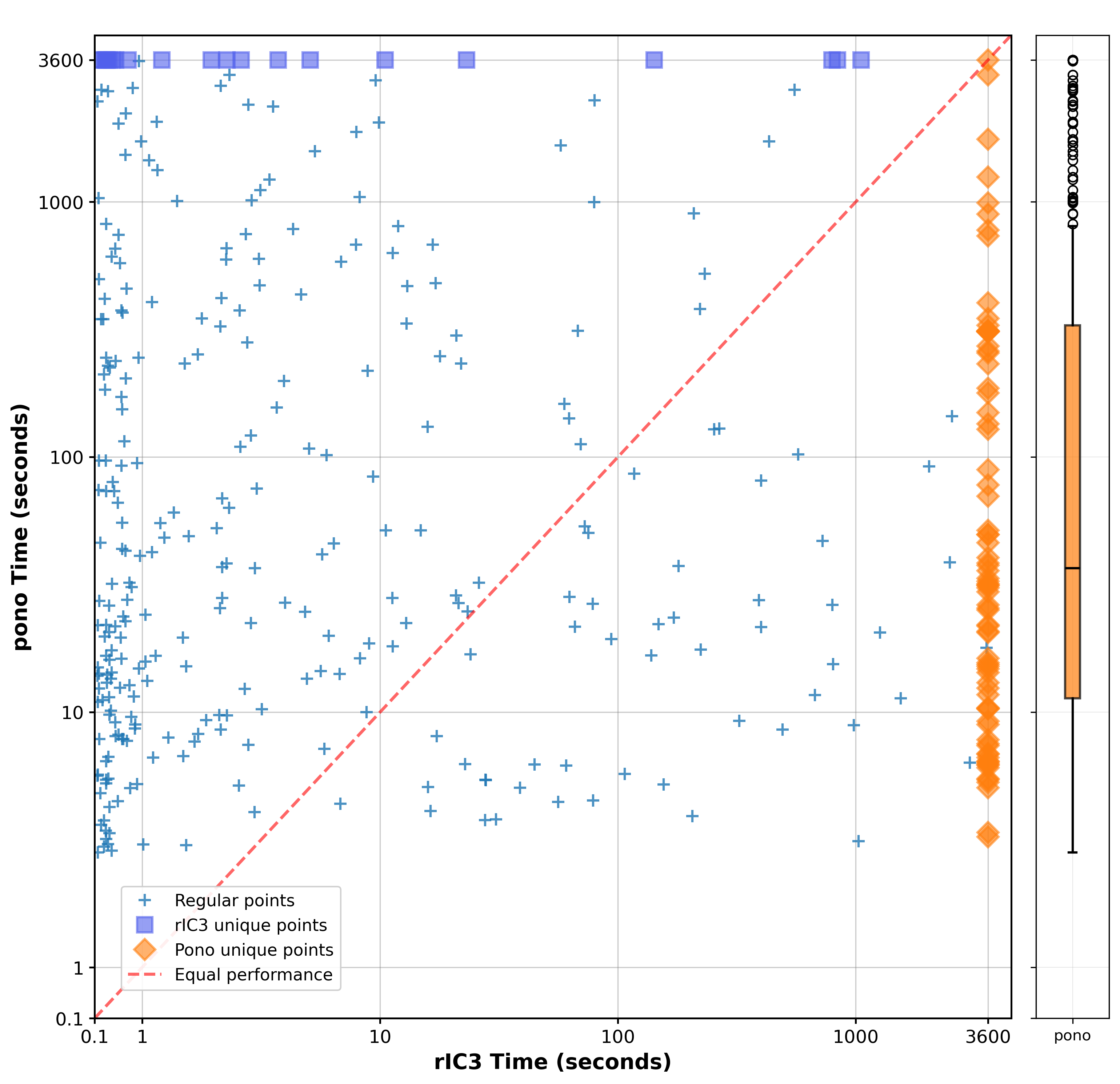}
    \caption{Performance comparison between rIC3 and Pono under the generated benchmark suite.}
    \label{fig:ex3_cmp}
\end{figure}

A key goal of a benchmark generator is to create instances that can clearly distinguish the performance of different solvers. We analyze whether the benchmarks generated by \OurMethod{} can reveal significant performance gaps between different state-of-the-art model checkers.
Fig.~\ref{fig:ex3_cmp} illustrates a performance comparison between two cutting-edge model checkers: rIC3 (uses dynamic solving) and Pono (uses IC3 with syntax-guided abstraction) on our generated benchmark suite in previous experiments. 
From the comparison, we can draw two conclusions: (1) Our benchmark suite distinguishes the performance of different model checkers. As can be observed from the figure, the distribution of points is far from the line representing equal performance. (2) Owing to the multi-agent algorithm that dynamically mutates and produces the computation graph with the guidance of Pono, the distribution of the Pono solving time exhibits a more balanced distribution compared with the solving time of the original HWMCC benchmark suite mentioned in Fig.~\ref{fig:motivation-a}. In the original HWMCC benchmark suite, the cases are either too simple or too hard to differentiate model checkers or provide meaningful evaluation for the performance of the model checker. According to the distribution on the right side of the figure, our \OurMethod can generate moderately difficult cases for the target model checker.  There is also an interesting observation: according to the result of HWMCC'24~\cite{biere2024hwmcc}, rIC3 beat Pono in the word-level track; with our benchmark suite, Pono outperforms the originally state-of-the-art rIC3.

\paragraph{Discussion of Current Limitations.}

This work targets model checkers in the PDR family because currently PDR-based model checkers are among the most powerful ones in recent HWMCC. In theory, \OurMethod{} can be extended to supporting algorithms like BMC and k-induction by extracting dynamic features from corresponding model checkers and re-training the predictor.  

The model checking problems generated by \OurMethod{} are functional equivalence checking problems between two hardware designs implementing the same high-level algorithm. This represents a common application scenario in hardware verification. 
The problem instances are not meant to completely replace existing HWMCC benchmarks but rather to offer a new perspective. 
\section{Conclusion}
\label{sec:conclusion}
%
%


This paper addressed the critical challenge of benchmark scarcity in the hardware model checking domain. We introduced \OurMethod{}, a novel framework that leverages reinforcement learning to guide the generation of benchmarks from a high-level algorithmic abstraction. By operating at the level of computation graphs and utilizing HLS to synthesize hardware, our approach systematically explores a rich space of structurally diverse and semantically complex designs.

Our experimental evaluation demonstrates that \OurMethod{} is effective in generating high-quality benchmarks, controllable in its ability to produce circuits with desired characteristics, and capable of revealing performance bottlenecks in state-of-the-art model checkers. The results confirm that our RL-guided approach is significantly more efficient at discovering challenging instances than random fuzzing strategies.

For future work, we plan to extend the action space to include a wider range of algorithmic and hardware-specific constructs, such as memory access patterns and interface protocols. Another promising direction is to explore multi-objective reward functions that consider not only solving time but also specific structural features of the generated circuits, further enhancing the controllability and targeted nature of the benchmark generation process.

\subsubsection*{Acknowledgements.}
This work is supported by the Guangdong S\&T Program (No. 2025A0505000022) and the Hong Kong Research Grants Council General Research Fund (GRF) (No. 16218324).

\subsubsection*{Code and Data Availability Statement.}
Our implementation, scripts, and documentation are available at:
\url{https://github.com/xfzhou01/EvolveGen}.
Reproducing the full experimental results requires access to AMD Xilinx Vitis HLS,
which is proprietary software and may require an appropriate license.
Users can obtain Vitis HLS through the vendor's official distribution and licensing channels.

\subsubsection*{Disclosure of Interests.}
The authors have no competing interests to declare that are relevant to the content of this article.



\newpage
\bibliography{refs}
\balance

\end{document}